\documentclass[letterpaper,keeplastbox,nospread  %a4paper
              ]{jacow}
\makeatletter%
	\ifboolexpr{bool{xetex}}
	 {\renewcommand{\Gin@extensions}{.pdf,%
	                    .png,.jpg,.bmp,.pict,.tif,.psd,.mac,.sga,.tga,.gif,%
	                    .eps,.ps,%
	                    }}{}
\makeatother

\ifboolexpr{bool{xetex} or bool{luatex}} % test for XeTeX/LuaTeX
 {}                                      % input encoding is utf8 by default
 {\usepackage[utf8]{inputenc}}           % switch to utf8

\usepackage[USenglish]{babel}			 

\usepackage[final]{pdfpages}
\usepackage{multirow}
\usepackage{ragged2e}

\ifboolexpr{bool{jacowbiblatex}}%
 {%
  \addbibresource{jacow-test.bib}
  \addbibresource{biblatex-examples.bib}
 }{}
\begin{document}

%\title{preparation OF papers for \NoCaseChange{JACoW} conferences\thanks{Work supported by ...}}
\title{Studies in Applying Machine Learning to LLRF and Resonance Control in Superconducting RF Cavities
    \thanks{The study at the University of New Mexico was supported by DOE Contract DE-SC0019468}}

\author{J. A. Diaz Cruz, S. G. Biedron, M. Martinez-Ramon, S. I. Sosa\\
 Department of Electrical and Computing Engineering\\
    University of New Mexico, Albuquerque, NM 87131, USA\\
    R. Pirayesh, Department of Mechanical Engineering,\\ University of New Mexico, Albuquerque, NM 87131, USA}
	
\maketitle
\begin{abstract}
Superconducting Radio-Frequency (SRF) cavities operating in continuous wave (CW) mode and with low beam loading are devices characterized by a high loaded quality factor, in the order of $10^7$, and narrow bandwidth, in the order of 10\,Hz. The Low Level RF (LLRF) and resonance control systems of SRF cavities become a fundamental component of the entire system operation and in general has very tight stability requirements on the amplitude, phase and resonance frequency of the cavity. Microphonics plays an important role in cavity detuning, which results in more power needed to achieve the desired gradient. Active control approaches to reduce detuning in SRF cavities using piezoelectric actuators have shown promising results. Furthermore, Machine Learning (ML) techniques have also shown important capabilities to improve existing PID controllers. Specifically, Neural Networks (NN) can be used to find optimal PID gains and improve performance of traditional control systems. In this research, we develop new approaches to improve existing LLRF and resonance control systems with ML as a tool to find the optimal gains. We investigate these approaches using the LLRF control system intended for LCLS-II.  
\end{abstract}

\section{Motivation}
Detuning due to microphonics can be reduced using both passive and active techniques. Passive techniques require a deep understanding of the resonant modes of the cavity and cryomodule environment, as well as identification of microphonics sources. Mechanical work should be performed to isolate and damp the source to decrease the coupling with the cavity and cryomodule. Previous work at Jefferson Lab and Fermilab has shown the benefits of this technique \cite{TPowers, Holzbauer}. In active techniques, piezoelectric tuners are used to compensate for microphonics detuning. In \cite{Holzbauer2}, a bank of narrow band filters was used for piezo tuning. In \cite{Banerjee}, a narrow band active noise control algorithm was demonstrated and an adaptive tuning approach was used to further tune control parameters.

ML is already a mature field with a solid theoretical background and with applications in several areas of science. The always increasing computational capabilities and resources have made possible to apply ML for image processing, self-driving cars, data-mining, among others, and it is not surprising that it has found applications in particle accelerators. In \cite{Edelen}, a Neural Network (NN) based control system was used to improve the capabilities of an existing PI resonance control system for an electron gun at Fermilab, reducing the settling time and overshoot of the temperature, which directly affects the resonance frequency. Back in 2007, a NN base self tuning PI feedback control system was developed for the LANSCE LLRF analog system at Los Alamos \cite{Kwon}. In this paper, we show progress from previous work \cite{Diaz} in using NNs as a tool to improve LLRF and Resonance control systems in SRF cavities.  

\section{LLRF and cavity model}
\begin{figure}[!hbt]
\centering
\includegraphics[height=3.5cm]{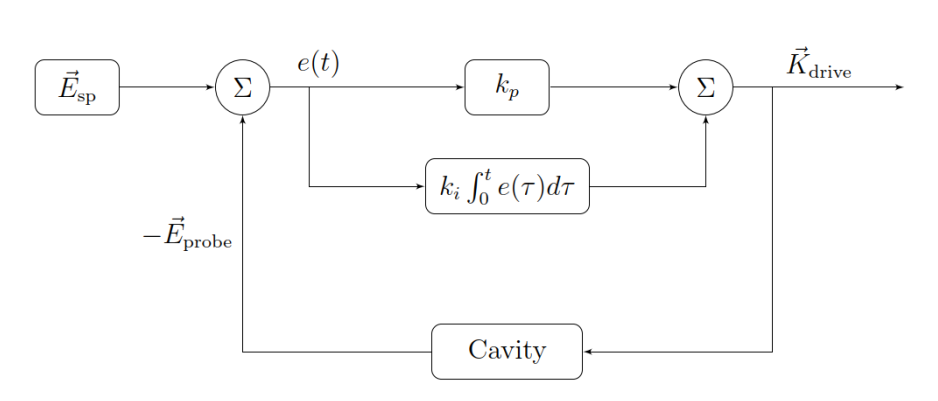}
\caption{Simplified diagram of a PI Controller for the LCLS-II LLRF System.}
\label{PI}
\end{figure}

In this research, a LLRF LCLS-II control system is used to model the SRF cavities. The LCLS-II is an upgrade to superconducting technology of the existing normal conducting linac at SLAC. The stability requirements for field amplitude and phase are 0.01\% and 0.01\,deg, respectively \cite{lcls2_llrf}. A Single Source Single Cavity (SSSC) topology was selected, with one power source (a Solid State Amplifier SSA) per superconducting cavity. The control system has been designed and tested, and it is in production phase. A simplified diagram is shown in Fig.\,\ref{PI}, where we can see how the system is based on a traditional PI controller \cite{cmoc}.

A nominal configuration has been chosen for the control system using the proportional gain, $k_p$, to increase the gain of the closed loop and the integral gain, $k_i$, to minimize the steady state error. The values are $k_p$ = 1200 and $k_i$ = 3.8$\times 10^7$ \cite{Serrano}. The purpose of this research is to dynamically tune these gains using ML to minimize error and energy. 

\begin{figure}[!hbt]
\centering
\includegraphics[height=4.0cm]{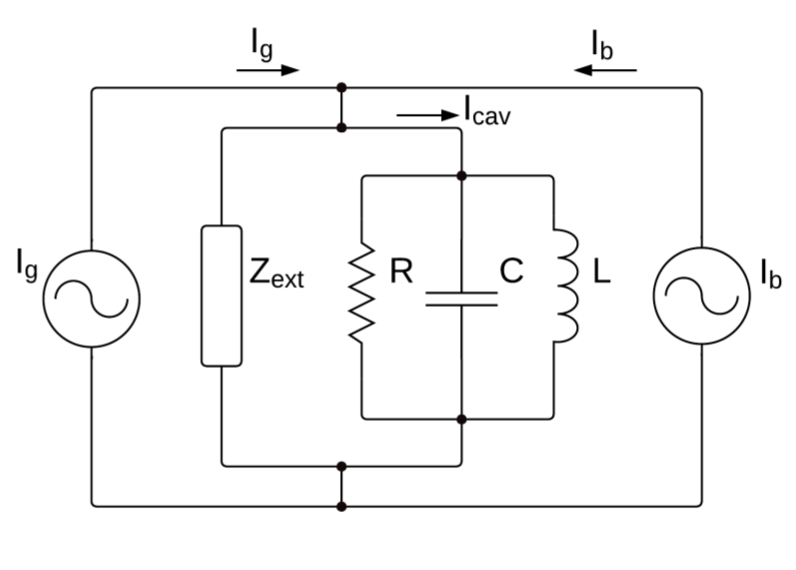}
\caption{Circuit model of a resonant mode in a cavity.}
\label{RLC}
\end{figure}

A model of the accelerator was developed by the LLRF team at LBNL, it is called \textit{Cryomodule-on-Chip} (CMOC) \cite{cmoc}, and it contains models of each resonant mode of an SRF cavity, RF stations, cryomodules and Linac sections. Each SRF cavity is modeled as a group of resonant modes, which are represented by the circuit model shown in Fig.\,\ref{RLC}.  

The equations that describe the electrodynamics of the system are explained in detail in \cite{cmoc}, and can be summarized in the following set of differential equations:
\begin{align}
V&=Se^{j\theta},\\
\frac{d\theta}{dt}&=w_d,\\
\frac{dS}{dt}&=-w_fS+w_fe^{-j\theta}(2K_g\sqrt{R_g}-R_bI),
\end{align}
where $V$ is a representative measure of each mode's energy, with magnitude $S$ and phase $\theta$, $w_d$ is the detuning frequency, which is affected by microphonics, $w_f$ is the cavity bandwidth, usually in the order of tens of hertz for superconducting cavities. $K_g$ is the incident wave amplitude, which represents the driven power coming from the SSA. $R_g$ is the coupling impedance of the beam, and $I$ represents the beam current. With this model, it is easy to simulate the cavity under the influence of only RF power, or both RF power and electron beam.

\section{Simulations}
\begin{figure}[!hbt]
\centering
\includegraphics[height=3.6cm]{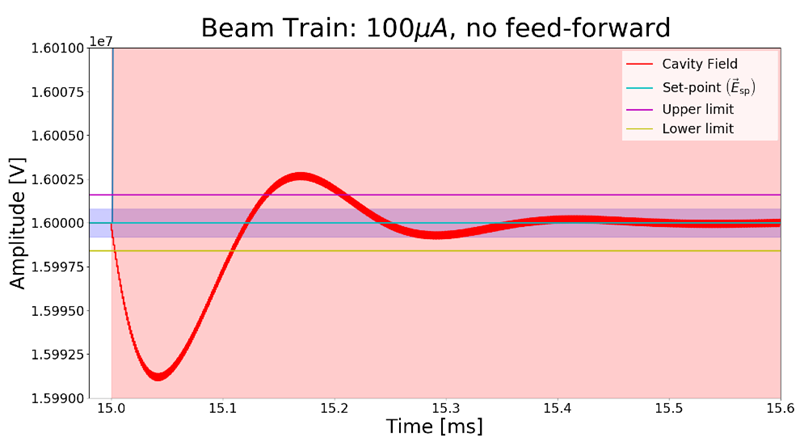}
\caption{Beam loading disturbance without feed-forward.}
\label{sim1}
\end{figure}

\begin{figure}[!hbt]
\centering
\includegraphics[height=3.6cm]{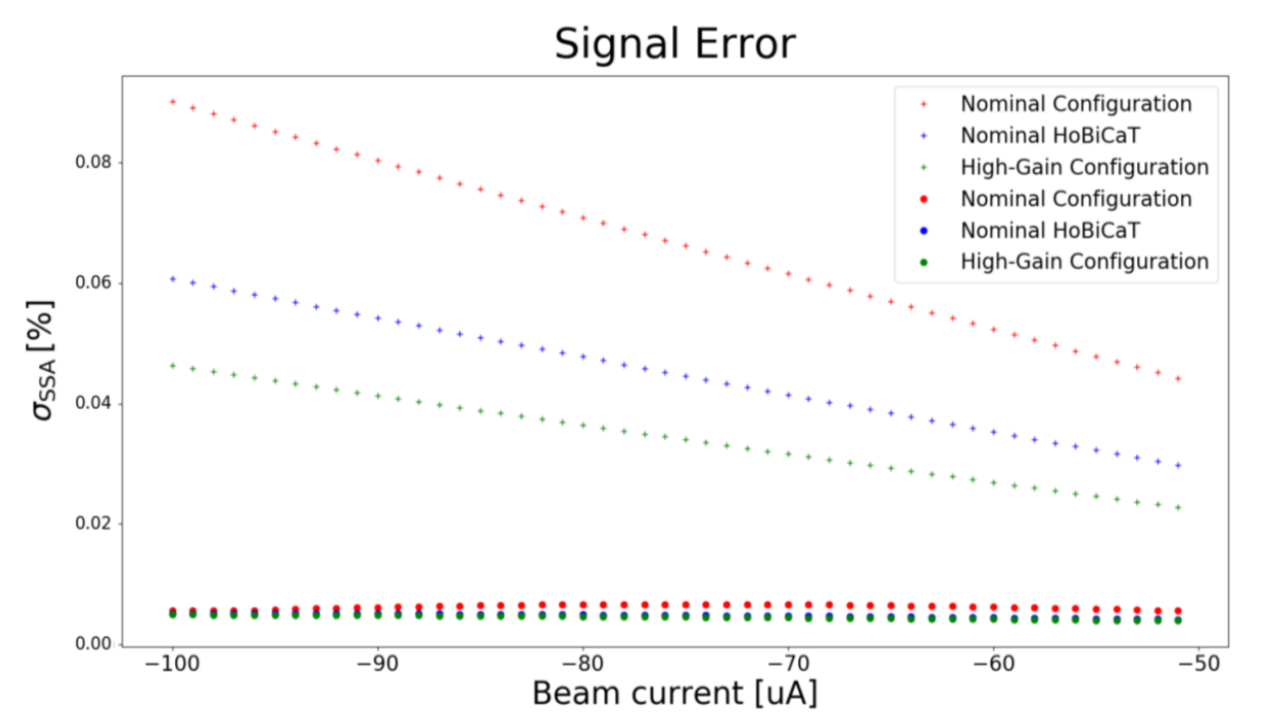}
\caption{Beam loading disturbance}
\label{sim2}
\end{figure}

Several types of perturbations can be simulated using the CMOC software tool. Beam loading disturbances, cavity constant detuning and measurement noise due to ADC and cables can be simulated and their influence in amplitude error can be characterized and studied. We have simulated all three sources of disturbances at different levels and under three gain configurations (nominal, HoBiCaT and high-gain configurations \cite{Serrano}). In Fig.\,\ref{sim1} we can see how the beam affects the amplitude stability. Levels of beam current varying from -100 to -50\,$\mu$A were used to generate the data plotted in Fig.\,\ref{sim2}. It is shown that for the higher currents, the stability of the amplitude decreases.

\begin{figure}[!hbt]
\centering
\includegraphics[height=3.6cm]{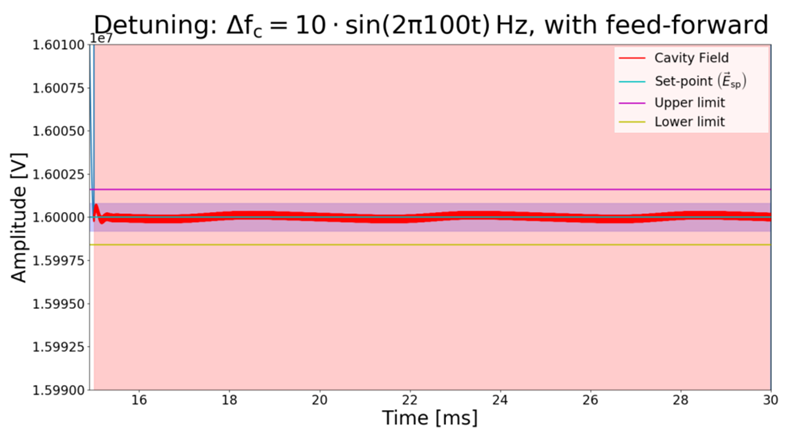}
\caption{Detuning with feed-forward.}
\label{sim3}
\end{figure}

\begin{figure}[!hbt]
\centering
\includegraphics[height=3.6cm]{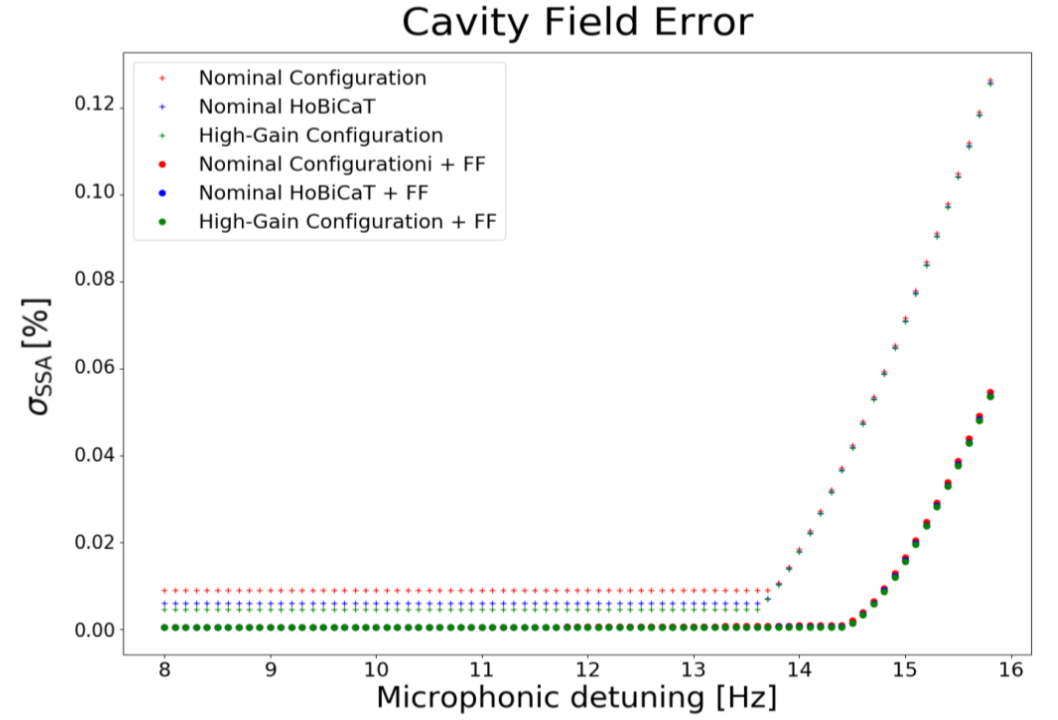}
\caption{Detuning impact on cavity field.}
\label{sim4}
\end{figure}

\begin{figure}[!hbt]
\centering
\includegraphics[height=3.6cm]{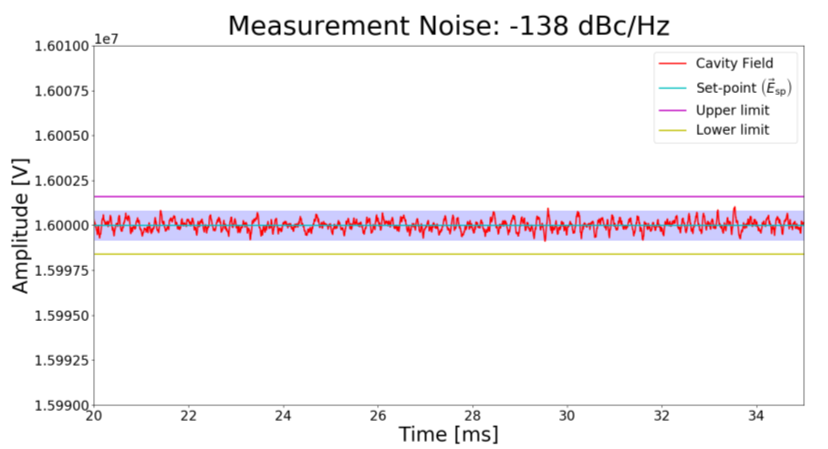}
\caption{Measurment noise with feed-forward.}
\label{sim5}
\end{figure}

\begin{figure}[!hbt]
\centering
\includegraphics[height=3.6cm]{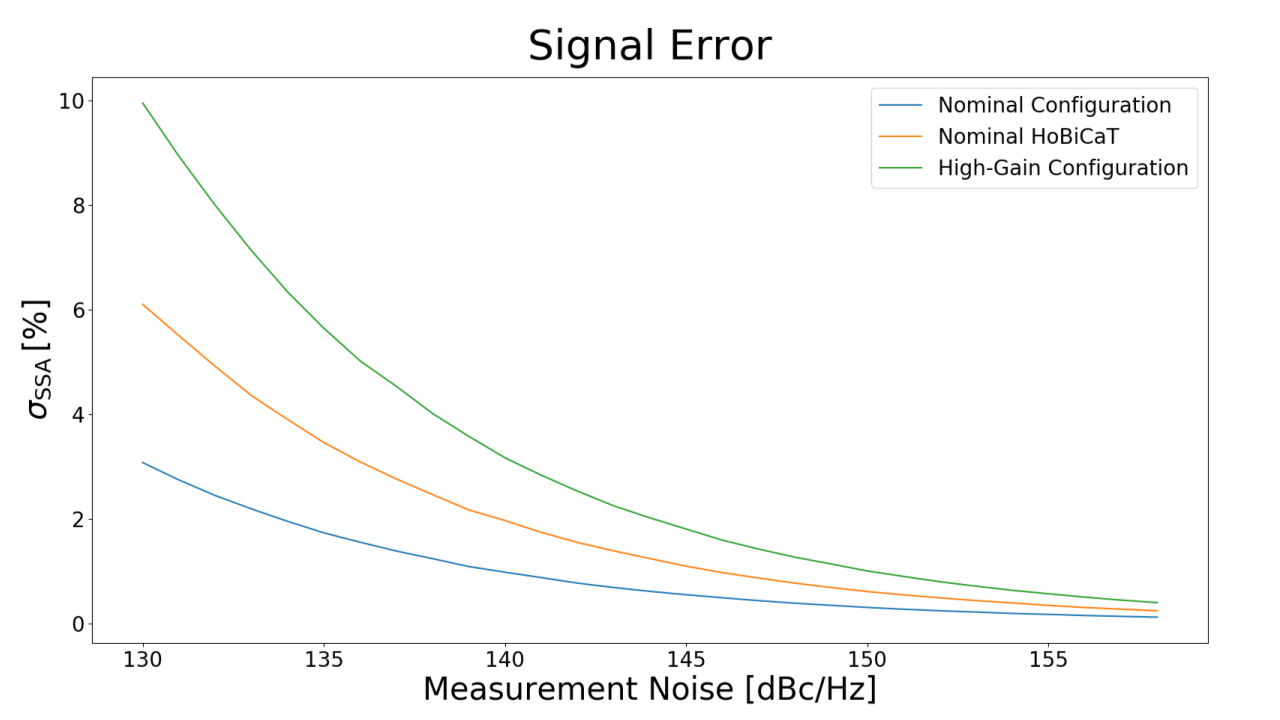}
\caption{Measurement noise impact on cavity field.}
\label{sim6}
\end{figure}

For constant microphonics detuning, Fig.\,\ref{sim3} shows the impact on amplitude with 10\,Hz detuning. Figure\,\ref{sim4} shows the data simulated under microphonics detuning from 8 to 16\,Hz. Finally, Fig.\,\ref{sim5} shows how the measurement noise affects the amplitude of the drive signal at -138\,dBc/Hz. Figure\,\ref{sim6} shows the data generated for measurement noise from -130 to -155\,dBc/Hz. With these simulations, it is illustrated that an optimal point needs to be found that minimizes error due to several disturbances. The data generated with these simulations will be used to train the proposed NNs. 


\section{AI framework design}
Complexity refers to the underlying information in a system. Particle accelerators are complex systems with numerous variables. Artificial intelligence (AI) can understand this complexity and use that knowledge to benefit the particle accelerators with high precision control systems, low maintenance, low costs, and high reliability operation. Resonance control of an SRF cavity includes several resonant modes and the model of the mechanical couplings between microphonics sources and cavity/cryomodules is not trivial; therefore, this research aims to increase the efficiency of the controller by processing the underlying complexity of the system using ML. 

\begin{figure}[!hbt]
\centering
\includegraphics[height=5.5cm]{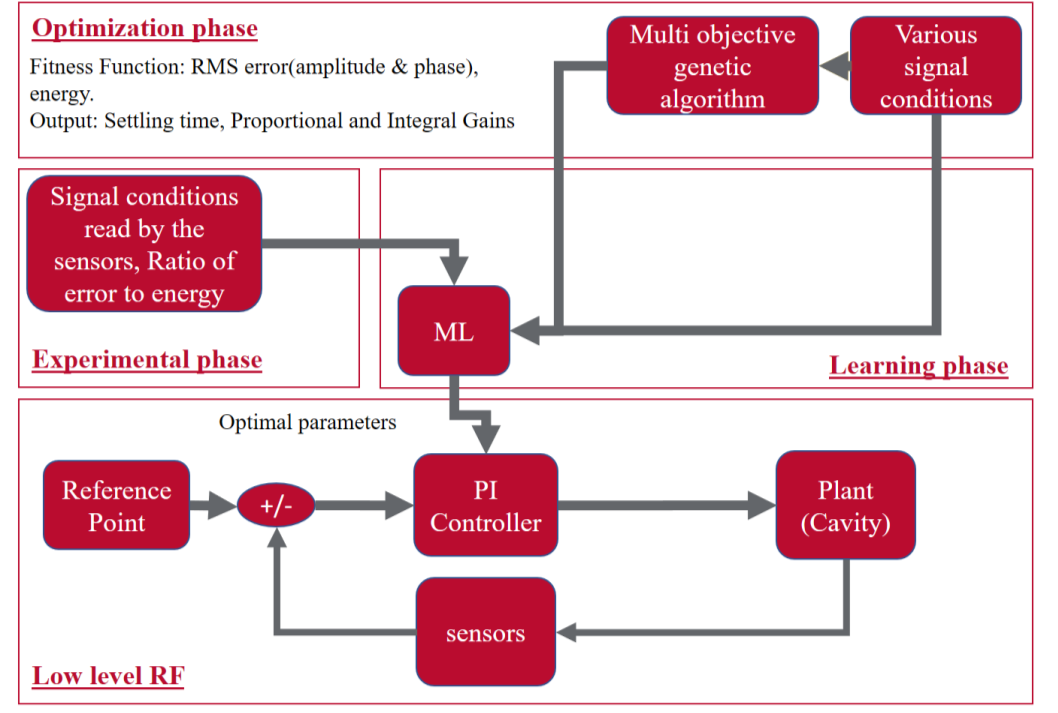}
\caption{Artificial Intelligence Framework.}
\label{AI}
\end{figure}

The variables that play role in the optimization phase are the cavity’s probe amplitude and phase and their rates of change (signal conditions in Figure \,\ref{AI}), the detuning, the RMS error of the stability, the RF output power, and the controller parameters (proportional and integral gains). In this control system, the energy and RMS error are being minimized with an optimization algorithm, and the other parameters are found to impose an optimum constraint on the control system. Figure \,\ref{AI} shows the hierarchy of AI that embeds the optimum set of parameters in the LLRF control system.

In the first phase of finding the optimal controller parameters, the optimization algorithm minimizes the loss function which is RMS error and energy consumption for each variable in the control system. As a result, the optimization, for each signal condition and detuning, produces an optimal controller parameter by minimizing the energy consumption and RMS error of the control system. There should be enough data produced by optimization for each signal condition and detuning. This phase is so crucial to the AI since this phase determines the accuracy of the estimation of the ML phase. To make the estimation of the ML more accurate, the data should be processed, cleaned, and analyzed, and enough data should be produced. The Pareto front figure can show how the error and energy changes with respect to each other. The more energy can cause less error and vice versa.

In the second phase, the data goes to ML algorithms in the training phase to produce functions that estimate the optimal controller parameters. In the experimental phase, the signal condition, the detuning, the energy consumption, and the RMS error define the optimal controller parameters. Finally, these optimal parameters are sent to the PI feedback loop. In a similar approach, we have developed \cite{reza1, reza2, reza3} and optimized \cite{reza4, reza5, reza6} the performance of control algorithms to formation control on satellites.

 \section{Summary and future work}
Microphonics affects the resonant frequency of SRF cavities. The detuning induced by microphonics increases the power consumption to maintain a specific gradient. Current passive and active control techniques can be used to reduce microphonics detuning. Here we presented an approach to improve amplitude and phase control by using AI as a tool to optimally tune the PI gains in a LLRF control system. Future work will be focused on applying these techniques to microphonics compensation with NNs.  

\section{Acknowledgements}
This work is being possible thanks to the guidance of the LCLS-II LLRf team, specially Larry Doolittle, Carlos Serrano and Gang Huang at LBNL, and Andy Benwell and Alex Ratti at SLAC.

\iffalse  % only for "biblatex"
	\newpage
	\printbibliography

% "biblatex" is not used, go the "manual" way
\else
%%\begin{thebibliography}{99}   % Use for  10-99  references

\null
\fi
\end{document}